\documentclass[amstex, preprint, prl, aps]{revtex4}

\usepackage{amsmath}

\begin{document}
 
\title{A quantum violation of the second law?}
\author{G. W. Ford}
\affiliation{Department of Physics, University of Michigan. Ann Arbor, MI 48109-1040}
\author{R. F. O'Connell}
\affiliation{Department of Physics and Astronomy, Louisiana State University, Baton
Rouge, LA 70803-4001}
\date{\today}
 
\begin{abstract}
An apparent violation of the second law of thermodynamics occurs when an
atom coupled to a zero-temperature bath, being necessarily in an excited
state, is used to extract work from the bath. Here the fallacy is that it
takes work to couple the atom to the bath and this work must exceed that
obtained from the atom. For the example of an oscillator coupled to a bath
described by the single relaxation time model, the mean oscillator energy
and the minimum work required to couple the oscillator to the bath are both
calculated explicitly and in closed form. It is shown that the minimum work
always exceeds the mean oscillator energy, so there is no violation of the
second law.
\end{abstract}
 
\maketitle
 
\section{Introduction}
 
Widespread interest in recent years in mesoscopic systems, fundamental
quantum physics and quantum computation has highlighted the critical role
which dissipative environments play in such studies. This has led to a
critical examination of many results that were derived for macroscopic
systems. In particular, there has been considerable interest in the area of
quantum and mesoscopic thermodynamics. In some instances questions have been
raised about the validity of the fundamental laws of thermodynamics,
especially at low temperatures where quantum effects are important \cite
{frontiers}. Whereas many interesting new facets of old results have
emerged, it is important to exercise caution before questioning the validity
of fundamental laws (especially the laws of thermodynamics), since many
subtle issues arise. Here we examine an apparent violation of the second law
of thermodynamics that would lead to the extraction of work from the zero
point fluctuations of the heat bath.
 
An atom in contact with a zero temperature heat bath is necessarily in an
excited state. This perhaps counter-intuitive observation is a simple
consequence of the fact that the Hamiltonian $H_{A}$ of the isolated atom
does not commute with the Hamiltonian $H$ of the \emph{system} of atom
coupled to the bath. It follows that in the ground state of the system the
energy of the atom \emph{must} fluctuate and therefore have a mean value
greater than its ground state energy $E_{0}$ \cite{ford95,buttiker02},
\begin{equation}
\left\langle H_{\text{A}}\right\rangle _{\mathrm{T=0}}>E_{0}.  \label{1.1}
\end{equation}
The existence of this energy might lead one to think that there could be a
quantum violation of the second law of thermodynamics, perhaps small, but
nevertheless a violation. Here the idea is to begin with the atom in its
ground state. It is then dropped into a bath at zero temperature, where it
comes to equilibrium with mean energy $\left\langle H_{A}\right\rangle _{
\mathrm{T=0}}$. The atom is then fished out and the energy of transition to
the ground state is used to raise a weight. Behold, we have a cyclic process
in which a weight is raised with no other effect than the extraction of
energy from a reservoir. This would be a violation of the second law in its
Kelvin-Planck form: \textquotedblright It is impossible to construct an
engine which will work in a complete cycle, and produce no effect excepting
the raising of a weight and the cooling of a heat
reservoir\textquotedblright\ \cite{planck}. What is wrong? Because there
must be something wrong. The answer is that it takes work to couple the atom
to the bath and this work exceeds that obtained in the transition to the
ground state. The principle of minimum work tells us that the minimum work
required to take a system from one thermodynamic state to another at the
same temperature is the difference of Helmholtz free energy \cite{landau_sp}
. In the present case this would be the free energy of the system of the
atom coupled to the bath minus the free energy of the bath in the absence of
the atom. \ One should expect, therefore, that if the second law is to hold
this work must be greater than the mean energy of the atom.
 
Our point here is that for the example of an oscillator coupled to a linear
passive heat bath both the mean energy and the work required to couple the
oscillator to the bath can be calculated explicitly and in closed form. With
the resulting expressions we are able to show that, for all coupling
strengths, the work is always greater than the mean energy and therefore
second law is satisfied.
 
\section{Oscillator coupled to the bath}
 
The mean energy of an oscillator coupled to a heat bath is given by the
expression
\begin{equation}
\left\langle H_{\text{O}}\right\rangle =\frac{1}{2}m\left\langle \dot{x}
^{2}\right\rangle +\frac{1}{2}K\left\langle x^{2}\right\rangle ,  \label{2.1}
\end{equation}
where $K$ is the oscillator force constant. To evaluate these expectations
we begin with the well known formula for the correlation function \cite
{ford88a},
\begin{equation}
\frac{1}{2}\left\langle x(t_{1})x(t_{2})+x(t_{2})x(t_{1})\right\rangle =
\frac{\hbar }{\pi }\int_{0}^{\infty }d\omega \coth \frac{\hbar \omega }{2kT}
\mathrm{Im}\{\alpha (\omega )\}\cos \omega (t_{2}-t_{1}).  \label{2.2}
\end{equation}
Here $\alpha (\omega )$ is the response function, of the general form
\begin{equation}
\alpha (\omega )=\frac{1}{-m\omega ^{2}-i\omega \tilde{\mu}(\omega )+K},
\label{2.3}
\end{equation}
in which $\tilde{\mu}(\omega )$ is the Fourier transform of the memory
function. For the single relaxation time model, this is given by
\begin{equation}
\tilde{\mu}(\omega )=\frac{\zeta }{1-i\omega \tau },  \label{2.4}
\end{equation}
in which $\zeta $ is the Ohmic friction constant and $\tau \ll m/\zeta $ is
the bath relaxation time. With this form of $\tilde{\mu}(\omega )$ the
response function is a rational function, with the denominator a cubic
polynomial in $\omega $. In order to perform the integrals we need it is
necessary to factor this denominator. For this purpose it is convenient to
introduce in place of the parameters $K$, $\zeta $ and $\tau $ the
parameters $\Omega $, $\omega _{0}$ and $\gamma $ through the relations
\begin{equation}
K=m\omega _{0}^{2}\frac{\Omega }{\Omega +\gamma },\quad \zeta =m\gamma \frac{
\Omega (\Omega +\gamma )+\omega _{0}^{2}}{(\Omega +\gamma )^{2}},\quad \tau =
\frac{1}{\Omega +\gamma }.  \label{2.5}
\end{equation}
With this replacement, the response function (\ref{2.3}) takes the form:
\begin{equation}
\alpha (\omega )=\frac{\omega +i(\Omega +z_{1}+z_{2})}{-m(\omega +i\Omega
)(\omega +iz_{1})(\omega +iz_{2})},  \label{2.6}
\end{equation}
where
\begin{equation}
z_{1}=\frac{\gamma }{2}+i\omega _{1},\quad z_{2}=\frac{\gamma }{2}-i\omega
_{1,}\quad \omega _{1}=\sqrt{\omega _{0}^{2}-\frac{\gamma ^{2}}{4}}.
\label{2.7}
\end{equation}
The poles of the response function are therefore at $\omega =-i\Omega $, $
-iz_{1}$, and $-iz_{2}$, all in the lower half plane. With this we see that
the relations (\ref{2.5}) are expressions for the physical parameters $K$, $
\zeta $ and $\tau $ in terms of the positions of the poles. The inversion of
these relations involves tedious expressions for the roots of a cubic
equation. However, the region of physical interest is that of $\tau $ small,
where $\Omega \sim 1/\tau $ is large, while $\omega _{0}^{2}\cong K/m$ and $
\gamma \cong \zeta /m$. Therefore, in the following expressions we use the
parameters $\Omega $, $\omega _{0}$ and $\gamma $, keeping in mind that $
\Omega $ is essentially the inverse of the bath relaxation time, while $
\omega _{0}$ and $\gamma $ are the oscillator natural frequency and
relaxation rate as slightly shifted by the coupling to the bath.
 
Setting $T=0$ in the formula (\ref{2.2}) and using the expression (\ref{2.6}
) for the response function, the integrals are elementary and we obtain the
exact expressions,
\begin{align}
\left\langle x^{2}\right\rangle _{\mathrm{T=0}}& =\hbar \frac{(\Omega
^{2}+\omega _{0}^{2}-\frac{\gamma ^{2}}{2})\arccos \frac{\gamma }{2\omega
_{0}}-\gamma \omega _{1}\log \frac{\Omega }{\omega _{0}}}{\pi m\omega
_{1}(\Omega ^{2}-\gamma \Omega +\omega _{0}^{2})},  \notag \\
\left\langle \dot{x}^{2}\right\rangle _{\mathrm{T=0}}& =\hbar \frac{\lbrack
\Omega ^{2}(\omega _{0}^{2}-\frac{\gamma ^{2}}{2})+\omega _{0}^{4}]\arccos
\frac{\gamma }{2\omega _{0}}+\gamma \omega _{1}\Omega ^{2}\log \frac{\Omega
}{\omega _{0}}}{\pi m\omega _{1}(\Omega ^{2}-\gamma \Omega +\omega _{0}^{2})}
.  \label{2.8}
\end{align}
With this we find from (\ref{2.1}) that the mean energy of the oscillator
coupled to the heat bath at temperature zero is given by
\begin{align}
\left\langle H_{\text{O}}\right\rangle _{\mathrm{T=0}}& =\frac{\hbar }{2\pi }
\left[ \frac{(\Omega ^{2}+\omega _{0}^{2})(2\Omega \omega _{1}^{2}+\gamma
\omega _{0}^{2})-\frac{1}{2}\gamma ^{3}\Omega ^{2}}{\omega _{1}(\Omega
+\gamma )(\Omega ^{2}-\gamma \Omega +\omega _{0}^{2})}\arccos \frac{\gamma }{
2\omega _{0}}\right.   \notag \\
& \qquad \left. +\frac{\gamma \Omega (\Omega ^{2}+\gamma \Omega -\omega
_{0}^{2})}{(\Omega +\gamma )(\Omega ^{2}-\gamma \Omega +\omega _{0}^{2})}
\log \frac{\Omega }{\omega _{0}}\right] .  \label{2.9}
\end{align}
The oscillator ground state energy is given by
\begin{equation}
E_{0}=\frac{\hbar }{2}\sqrt{\frac{K}{m}}=\frac{\hbar \omega _{0}}{2}\sqrt{
\frac{\Omega }{\Omega +\gamma }}.  \label{2.10}
\end{equation}
As we have remarked, the mean energy must always be greater than the ground
state energy. This is clearly seen in Fig 1 where a
three-dimensional plot of $\left\langle H_{\text{O}}\right\rangle _{\mathrm{
T=0}}/E_{0}$ is shown for a wide range of $\gamma /\omega _{0}$ and $\Omega
/\omega _{0}$.
 
Next we consider the\ work required to couple the oscillator to the bath. \
The principle of minimum work tells us that the minimum work required to
take a system from one thermodynamic state to another is the difference of
Helmholtz free energy \cite{landau_sp}. In the present case this would be
the free energy of the system of the oscillator coupled to the bath minus
the free energy of the bath in the absence of the oscillator. For this free
energy difference there is the remarkable formula \ which for the system at
temperature $T$ takes the form \cite{ford85}:
\begin{equation}
F_{\text{O}}(T)={\frac{1}{\pi }}\int_{0}^{\infty }d\omega f(\omega ,T)
\mathrm{Im}\{{\frac{d\log \alpha (\omega +i0^{+})}{d\omega }}\},
\label{2.11}
\end{equation}
where $\alpha (\omega )$ is again the oscillator susceptibility. In this
formula $f(\omega ,T)$ is the free energy (including zero point energy) of a
free oscillator of frequency $\omega $ at temperature $T$:
\begin{equation}
f(\omega ,T)=kT\log (2\sinh \frac{\hbar \omega }{2kT}).  \label{2.12}
\end{equation}
At zero temperature $f(\omega ,0)=\hbar \omega /2$ and using again the
expression (\ref{2.6}) for the response function, the integral (\ref{2.11})
can be evaluated exactly to give the simple result:
\begin{equation}
F_{\text{O}}(0)=\frac{\hbar }{2\pi }\left[ 2\omega _{1}\arccos \frac{\gamma
}{2\omega _{0}}+\gamma \log \frac{\Omega }{\omega _{0}}+(\Omega +\gamma
)\log \frac{\Omega +\gamma }{\Omega }.\right] .  \label{2.13}
\end{equation}
 
In Fig. 2 we plot $F_{\text{O}}(0)/E_{0}$ and $\left\langle
H_{O}\right\rangle _{\mathrm{T=0}}/E_{0}$ vs the coupling strength as
measured by $\gamma /\omega _{0}$. In this plot we have chosen $\Omega
=5\omega _{0}$ but for any $\Omega >\omega _{0}$ the result is the same: the
minimum work required to couple the oscillator to the bath, $F_{\text{O}}(0)$
, is always greater than the mean energy of the oscillator coupled to the
bath,$\left\langle H_{O}\right\rangle _{\mathrm{T=0}}$. Thus in the cyclic
process described above there is a loss rather than a gain of energy and
there is no violation of the second law.
 
\section{Remarks}
 
In the limit of vanishing bath relaxation time, both $F_{\text{O}}(0)$ and $
\left\langle H_{O}\right\rangle _{\mathrm{T=0}}$ are logarithmically
divergent, but it is not difficult to see that
\begin{equation}
F_{\text{O}}(0)-\left\langle H_{O}\right\rangle _{\mathrm{T=0}}\rightarrow
\frac{\gamma }{\pi \omega _{0}}E_{0}.  \label{3.1}
\end{equation}
From the figure it is clear that this linear dependence on $\gamma $ is a
pretty good approximation even for rather long bath relaxation times. The
discrepancy between the work required to couple the oscillator to the bath
and that obtainable from the oscillator removed from the bath increases
roughly proportional to the strength of coupling to the bath.
 
The free energy at zero temperature is in fact the energy. Therefore, we see
that coupling the oscillator to the bath actually requires an energy greater
than $\left\langle H_{O}\right\rangle _{\mathrm{T=0}}$, the additional
energy going to the bath. Since the bath is infinite, this finite additional
energy does not raise the temperature.
 
The principle of minimum work becomes the principle of maximum work when the
initial \ and final states are reversed \cite{landau_sp}. Returning to the
general case of an atom we considered in the beginning, if $F_{\mathrm{A}
}(T) $ is the free energy of the atom coupled to the bath at an arbitrary
temperature $T$ minus the free energy of the bath alone, then $F_{\mathrm{A}
}(T)$ is the minimum work required to couple the atom to the bath while it
is also the maximum work that can be obtained when the atom is extracted
from the bath. The net work obtained can never be positive.
 
\begin{acknowledgments}
Part of this work was carried out at the School of Theoretical Physics of
the Dublin Institute for Advanced Studies . The authors wish to thank the
School and its Director, Professor Tony Dorlas for their hospitality.
\end{acknowledgments}
 
\newif\ifabfull\abfulltrue

\newpage

\begin{center}
Figure Captions
\end{center}
 
Fig. 1. Three dimensonal plot of $\left\langle H_{\text{O}}\right\rangle _{%
\mathrm{T=0}}/E_{0}$ as a function of $\ \gamma /\omega _{0}$ and $\Omega
/\omega _{0}$.
 
Fig. 2. Plot of $F_{\text{O}}(0)/E_{0}$ and $\left\langle H_{\text{O}%
}\right\rangle _{T=0}/E_{0}$ as functions of the coupling strength as
measured by $\gamma /\omega _{0}$. The relaxation time is chosen so that $%
\Omega =5\omega _{0}$. The overdamped oscillator corresponds to $\gamma
/\omega _{0}>2$.

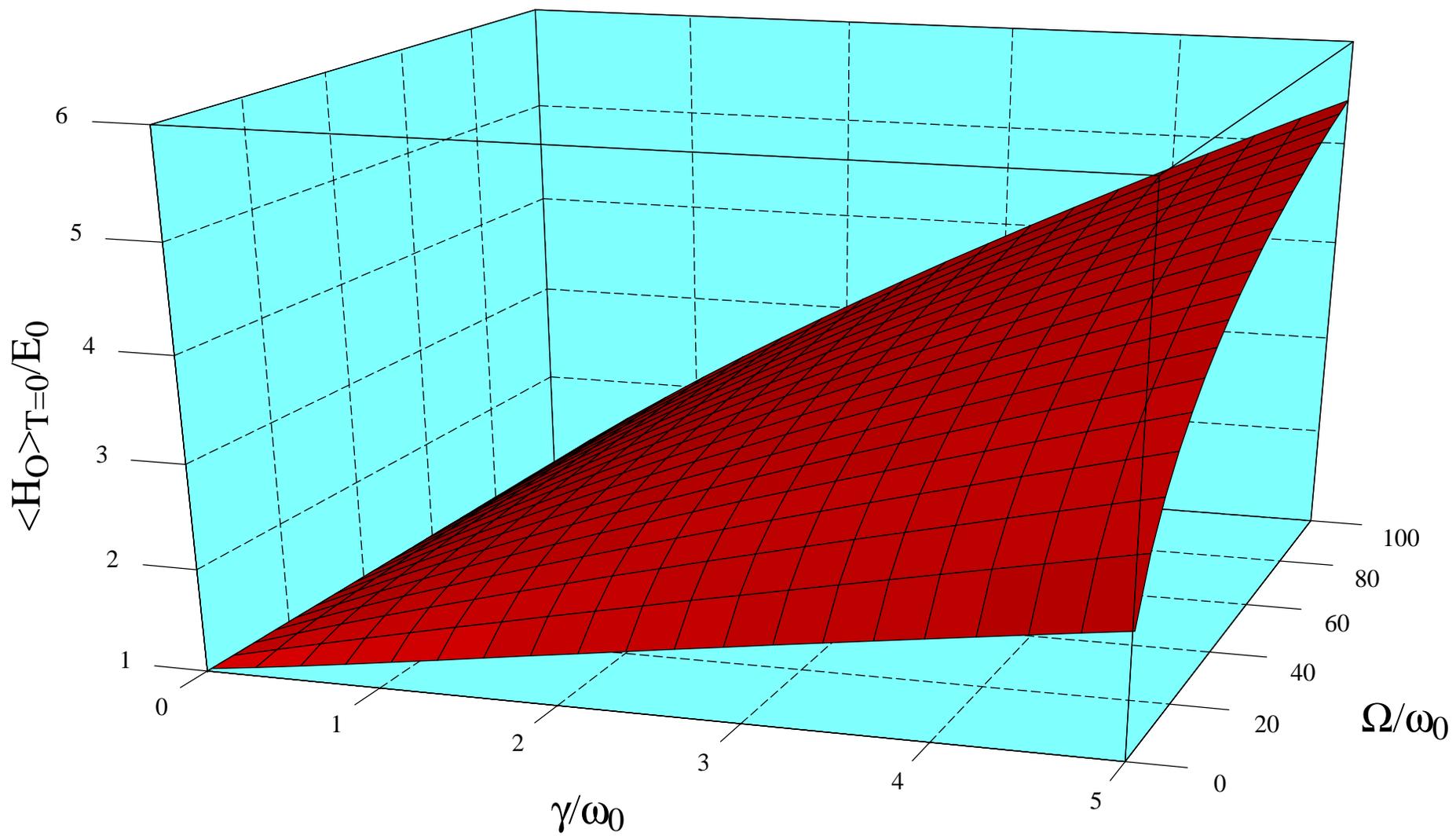
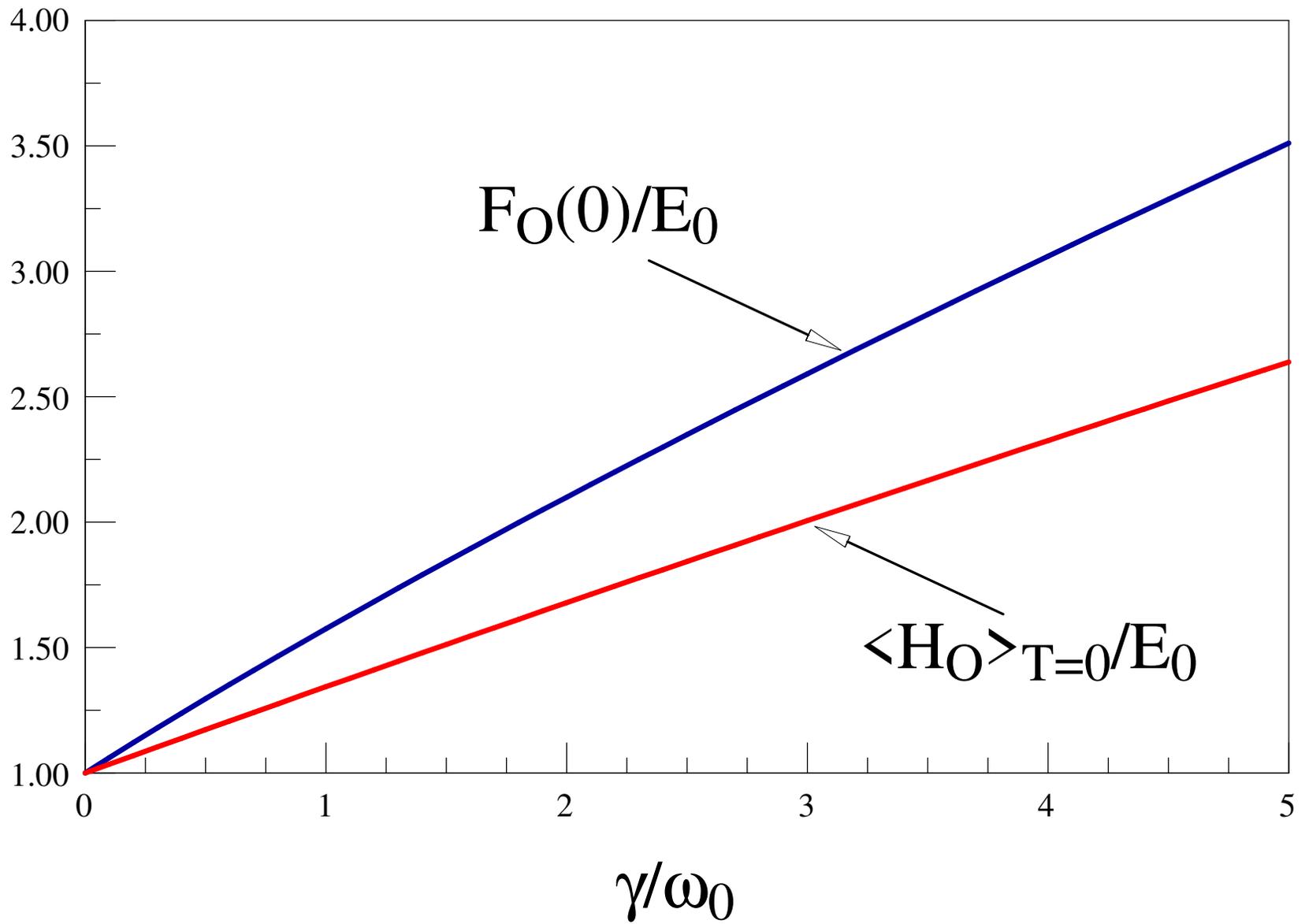
\end{document}